\documentstyle[12pt]{article}

\begin{document}

\date{}
\title{Classical Geometric Interaction- picture-like Description \thanks{%
this work is supported by NSF of China, Pan Den Plan of China and LWTZ -1298
of Chinese Academy of Sciences}}
\author{{\ Ming-Xue Shao}$^{123}${\thanks{{\protect\small
E-mail: shaomingxue@hotmail.com}} , Zhong-Yuan
Zhu}$^{23}${\thanks{{\protect\small E-mail: zzy@itp.ac.cn}}}\\
$^1${\small Department of Physics, Beijing Normal University, Beijing
100875, P.R.China.}\\
$^2${\small CCAST(World Laboratory), P.O.Box 8730, Beijing, 100080, 
P.R.China%
}\\ $^3${\small Institute of Theoretical Physics, Academia
Sinica, P.O.Box 2735, Beijing 100080, P.R.China.}}
\maketitle
\date{}

\begin{abstract}
In order to get the classical analogue of quantum interaction picture in
classical symplectic geometric description, the space of solutions of free
equations of motion is suggested to replace the phase space in $T^{*}Q$
description or the space of motions in usual classical symplectic geometric
description. The way to determine measured values of observables in this
scheme is worked out.\newline

Mathmatics Subject Classification(1991): 81S10 \newline
\end{abstract}

\vskip 0.6in

It is well known that there exist many different pictures in
quantum formalism. The most famous and useful ones are
Schr\"odinger, Heisenberg and interaction pictures. During recent
years a relatively new quantization formulation called geometric
quantization has been paid attentions and been considered as so
far mathematically most thorough approach to quantization. In
short, geometric quantization \cite{Wood}\cite{Sniatycki} is
essentially a globalization of canonical quantization. It is
heavily based on the symplectic geometrical description of
classical system. In books on the geometric quantization,
usually, ones start from a phase space (for example,
tangent bundle $TQ$ or cotangent bundle $T^{*}Q$ of configuration space $Q$%
). But, as pointed out in ref.\cite{Witten}, this description is
not obviously relativistically covariant in relativistic theory
since at the beginning a special time should be chosen. Thus,
another description, which uses the space of solutions of the
equation of motion (called space of motion $M$) as the state
space\cite{souriau}, has been used to establish covariant
geometric formalisms of various relativistic field theories by
Crnkovic and Witten et al.
\cite{Witten}\cite{Crnkovic}\cite{Zuckman}\cite {LSZ}. In fact,
as pointed in P.21 of the ref.\cite{Wood}, above two descriptions
are classical analogues of quantum Schr\"odinger and Heisenberg
pictures respectively. Thus a natural question arises. How to
establish a classical analogue of the interaction picture in
classical symplectic geometric description since the interaction
picture is very important in quantum perturbative calculation. In
this paper, we investigate this problem and work out it based on
the space of solutions of free equations of motion. We illustrate
its relationships with above two classical geometric descriptions
are similar to the relationships between the interaction,
Schr\"odinger and Heisenberg Pictures of quantum mechanics. Thus
in geometric descriptions the classical and quantum structures in
this respect are given in parallel .

Let us consider a conservative dynamical system for which the set of
kinematically possible states can be represented by a velocity phase space
or tangent bundle $TQ$. Its dynamical behavior is determined by a Lagrangian
$L(q^a,{\dot{q}}^a),$ which is regular in the following sense, i.e.,
\begin{equation}
\det (\frac{\partial ^2L}{\partial \stackrel{.}{q}^a\partial 
\stackrel{.}{q}%
^b})\neq 0    \label{0}
\end{equation}
everywhere. The Hamiltonian $h$ of the system can be defined by
\begin{equation}
h(q^a,{\dot{q}}^a)=\stackrel{.}{q^a}\frac{\partial L}{\partial 
\stackrel{.}{q%
}^a}-L .   \label{0a}
\end{equation}
We also require that the Hamiltonian vector field $X_h$ is complete. The
equations of motion are given by the Lagrangian equations
\begin{equation}
\frac d{dt}(\frac{\partial L}{\partial \stackrel{.}{q}^a})-\frac{\partial 
L}{%
\partial q^a}=0 .   \label{d0}
\end{equation}
In order to transfer to the momentum phase space(or cotangent bundle 
$T^{*}Q$
) ones take the Lengendre transformation
\begin{equation}
\rho :TQ\rightarrow TQ^{*}:(q^a,{\dot{q}}^a)\mapsto (q^a,p_a),
\label{b0}
\end{equation}
where locally
\begin{equation}
p_a=\frac{\partial L(q,{\dot{q}})}{\partial \stackrel{.}{q^a}}.
\label{c0}
\end{equation}
In this scheme Hamiltonian becomes $H(q,p)=h(q,{\dot{q}}(q,p)).$ The
equations of motion are given by the Hamiltonian equations,
\begin{eqnarray}
\stackrel{.}{q}^a(t)   & = &  \frac{\partial
H(q^a(t),p_a(t))}{\partial p_a(t)}, \label{e0} \\
\stackrel{.}{p}_a(t)   & =  & -\frac{\partial
H(q^a(t),p_a(t))}{\partial q^a(t)}. \nonumber
\end{eqnarray}

In order to discuss interaction- picture-like description let us write the
Lagrangian into two parts,
\begin{equation}
L=L_0+L_1,    \label{1}
\end{equation}
where $L_0$ is a free Lagrangian\footnote{%
In fact following approach is suitable for any regular $L_0$. In practice
ones often take a Lagrangian whose equation of motion has rigorous solutions
as $L_0$.} and $L_1$ is the interaction part. Thus the equation of free
motion of this system is given by

\begin{equation}
\frac d{dt}(\frac{\partial L_0(q_0^a(t),{\dot{q}}_0^a(t))}{\partial
\stackrel{.}{q_0}^a(t)})-\frac{\partial L_0(q_0^a(t),{\dot{q}}_0^a(t))}{%
\partial q_0^a(t)}=0.    \label{2}
\end{equation}
The manifold of solutions of the equation $(\ref{2})$ of free
motions $M_F$ can be defined as follows. From now on, $q_0$ will
be used to denote a free solution mapping, which is defined to be
\begin{equation}
q_0:t\mapsto (q_0^a(t)),    \label{2a}
\end{equation}
where $(q_0^a(t))$ satisfies the equation $(\ref{2})$. $M_F$
consists of all solution mappings of the eq.$(\ref{2}).$ It can
be given a topology and made into a manifold by using the values
of $q_0^a(t_0)$ and ${\dot{q}}_0^a(t_0)$ at some particular time
$t_0$ as coordinates. The mapping
\begin{eqnarray}
\tau _t    &:&   M_F\rightarrow TQ    \nonumber \\
   &:&   q_0\mapsto (q_0^a(t),{\dot{q}}_0^a(t))
\label{b2}
\end{eqnarray}
is a diffeomorphism since the free Lagrangian $L_0$ is regular and $X_{h_0}$
is complete. Note that if $q_0:t\mapsto (q_0^a(t))$ is a solution of 
(\ref{2}%
), so is $q_0^{a\prime }:t\mapsto (q_0^{a\prime }(t)\equiv q_0^a(t+k))$
where $k$ is a constant. Generally speaking, it is a distinct solution so
that regarded as a distinct point of $M_F,$ even though the two orbits
occupy the same point set in $Q$.

A tangent vector $U$ to $M_F$ at a solution $q_0$ is represented by a
solution $u_0$ of the linearized equations of free motion $(\ref{2})$%
\[
u_0:t\longmapsto u_0(t)
\]
and $u_0(t)$ satisfies

\begin{equation}
\left. \left\{ \frac d{dt}(\frac{\partial ^2L_0}{\partial \stackrel{.}{q_0}%
^a\partial \stackrel{.}{q}_0^b}\stackrel{.}{u_0^b}+\frac{\partial ^2L_0}{%
\partial \stackrel{.}{q_0}^a\partial q^b}u_0^b)-\frac{\partial ^2L_0}{%
\partial q_0^a\partial \stackrel{.}{q_0}^b}\stackrel{.}{u_0}^b-\frac{%
\partial ^2L_0}{\partial q_0^a\partial q_0^b}u_0^b\right\}
\right|_{ q_0^a=q_0^a(t)}^{u_0^b=u_0^b(t)} =0. \label{3}
\end{equation}
The free part of action is

\begin{equation}
I_0(t_2,t_1)=\int\limits_{t_1}^{t_2}dt\cdot L_0(q_0^a(t),{\dot q}_0^a(t)),
\label{r4}
\end{equation}
where $t_1$ and $t_2$ are fixed values of the time. $I_0(t_2,t_1)$ can be
considered as a function on $M_F$ when, as shown in (\ref{r4}), the
arguments of $L_0$ are restricted to solutions of the free equation. Thus
the derivative of $I_0(t_2,t_1)$ along $U$ is given by

\begin{eqnarray}
U\circ dI_0 &=& \int\limits_{t_1}^{t_2}dt\cdot \left[
\frac{\partial L_0(q_0^a(t),{\dot q}_0^a(t))}{\partial
q_0^a(t)}u_0^a(t)+\frac{\partial
L_0(q_0^a(t),{\dot q}_0^a(t))}{\partial \stackrel{.}{q}_0^a(t)}\stackrel{%
\cdot }{u}_0^a(t)\right]    \nonumber \\
\ & =&\left. \left[ \frac{\partial L_0}{\partial \stackrel{.}{q_0}^a(t)}%
u_0^a(t)\right] \right|
_{t_1}^{t_2}+\int\limits_{t_1}^{t_2}dt\;u_0^a(t)\left[
\frac{\partial L_0(q_0^a(t),{\dot q}_0^a(t))}{\partial
q_0^a(t)}-\frac d{dt}(\frac{\partial L_0(q_0^a(t),{\dot
q}_0^a(t))}{\partial \stackrel{.}{q_0}^a(t)})\right] . \label{5}
\end{eqnarray}
Now the final integral in (\ref{5}) vanishes because $q_0$ is a solution of
the equation ($\ref{2}$). Thus we find we can define for each $t$ a 1-form 
$%
\theta _t$ on $M_F$ by$_{}$

\begin{equation}
U\circ \theta _t=u_0^a(t)\frac{\partial L_0(q_0^a(t),{\dot q}_0^a(t))}{%
\partial \stackrel{.}{q}_0^a(t)},    \label{6}
\end{equation}
so that the equation ($\ref{5}$) implies that
\begin{equation}
dI_0(t_2,t_1)=\theta _{t_2}-\theta _{t_1}.    \label{7}
\end{equation}
Since $dI_0(t_2,t_1)$ is an exact form on $M_F,$ we get from (\ref{7}) that

\begin{equation}
d\theta _{t_2}-d\theta _{t_1}=0.    \label{8}
\end{equation}
So the closed form in terms of coordinates $(q_0^a(t),{\dot q}_0^a(t))$
\begin{equation}
\omega _T\equiv d\theta _t=d\frac{\partial L_0(q_0^a(t),{\dot q}_0^a(t))}{%
\partial \stackrel{.}{q}_0^a(t)}\wedge dq_0^a(t)    \label{a8}
\end{equation}
on $TQ$ does not depend on $t$ and, its pull-back $\tau _t^{*}\omega _T$
provides a natural symplectic structure on $M_F.$ From this symplectic
structure we can define Hamiltonian vector fields, Poisson bracket etc. on 
$%
M_F.$

Above presentation uses tangent bundle $TQ.$ In practical
application the momentum phase space (or cotangent bundle
$TQ^{*}$) is also often used.
Similarly to $\tau _t$ in (\ref{b2}), we can define another mapping $\Pi 
_t$%
\begin{eqnarray}
\Pi _t   & : &  M_F  \mapsto T^{*}Q    \nonumber \\
\       & : & q_0  \mapsto (q_0^a(t), p_{a0}(t)),    \label{b8}
\end{eqnarray}

i.e.
\begin{eqnarray}
(\stackrel{\wedge }{q}^a\circ \Pi _tq_0,\stackrel{\wedge
}{p}_a\circ \Pi _tq_0)   & = &  (q_0^a(t),p_{a0}(t))  \nonumber
\\ \   & \equiv  &  \Pi _tq_0 .   \label{c8}
\end{eqnarray}
In (\ref{b8}) $q_0$ is the same solution of free motion in (\ref{b2}).
Equivalently, instead of eq.$(\ref{2}),$ we have ($q_0^a(t),p_0^a(t)$)
satisfies the Hamiltonian equations of free motion
\begin{eqnarray}
\stackrel{.}{q}_0^a(t) &=& X_{H_0(q_0^a(t),p_{a0}(t))}\;q_0^a(t),
\nonumber \\ \stackrel{.}{p}_{a0}(t)
  & = &  X_{H_0(q_0^a(t), p_{a0}(t))}\;p_{a0}(t).
\label{9a}
\end{eqnarray}
If we take $(q_0^a(t_0),p_{a0}(t_0))$ at a given time $t_0$ as coordinates
of $M_F$, by using them we can write the symplectic 2-form as
\begin{equation}
\omega _c=dp_{a0}(t_0)\wedge dq_0^a(t_0).    \label{a10}
\end{equation}
It is easy to prove that $\omega _c$ does not depend on $t_0$ since the
transformation $(q_0^a(t_0),p_{0a}(t_0))\rightarrow (q_0^a(t),p_{0a}(t))$ is
a canonical transformation. Therefore its pullback $\Pi _{t_o}^{*}\omega _c$
is well defined on $M_F$ and provides a symplectic 2- forms. Comparing 
eqs.(%
\ref{a10}) and (\ref{a8}), we find $\Pi _{t_o}^{*}\omega _c=\tau
_t^{*}\omega .$

Now let us turn to discuss how to determine the classical measured values of
observables at time $t$ in this scheme. First of all we like to point out
that $H_1(q_0^a(t),p_{a0}(t))\in R$ depends on not only solution $q_0,$ but
also the time $t.$ So if we define a function $(\tau _t^FH_1)$on $M_F$ as
follows
\begin{eqnarray}
(\Pi _t^{*}H_1) &:&  M_F\rightarrow R    \nonumber \\
\ &:&  q_0\mapsto (\Pi _t^{*}H_1)q_0\equiv H_1(q_0(t),p_0(t)).
\label{10d}
\end{eqnarray}
it is $t-$dependent. From eqs. (\ref{9a}) and (\ref{10d}), the
$t-$ dependence of the quantity $(\Pi _t^{*}H_1)$ is determined
by the free Hamiltonian. Then from its Hamiltonian vector field
$X_{\Pi _t^{*}H_1}$ ones can generate a curve on $M_F$
\begin{equation}
\frac d{dt}q_0^{(t)}=X_{\Pi _t^{*}H_1}q_0^{(t)}    \label{10e}
\end{equation}
with $t$ as the parameter of the curve. Here $q_0^{(t)}$ represents the
state in such description$.$ Note since $\Pi _t^{*}H_1$ is dependent on
parameter $t$ , so does the generator $X_{\Pi _t^{*}H_1}$. Furthermore from
above settings we define
\begin{equation}
(q^a(t),p_a(t))\equiv (q_0^{a(t)}(t),p_{a0}^{(t)}(t))=\Pi
_tq_0^{(t)}=\left. \Pi _tq_{0~}^{(s)}\right|_{s=t}. \label{r11}
\end{equation}
We shall prove that $(q^a(t),p_a(t))$ defined here satisfies the
Hamiltonian equations (\ref{e0}). In fact,
\begin{eqnarray}
(\frac{dq^a(t)}{dt},\frac{dp_a(t)}{dt}) &=& \frac d{dt}(\Pi
_tq_0^{(t)})=\frac d{dt}(q^{a(t)}(t),p_a^{(t)}(t))
\nonumber
\\ \ & =& \left.
(\frac{dq_0^{a(s)}(t)}{dt},\frac{dp_{a0}^{(s)}(t)}{dt})\right|
_{s=t}+\left.
(\frac{dq_0^{a(s)}(t)}{ds},\frac{dp_{a0}^{(s)}(t)}{ds})\right|
_{s=t} \nonumber \\ \ & =& \left.
X_{H_0(q_{0(t)}^{a(s)},p_{a0}^{(s)}(t))}\Pi _tq_0^{(s)}\right|
_{s=t}+\left. \Pi _tX_{\Pi _t^{*}H_1(q_0^{(s)})}q_0^{(s)}\right|
_{s=t} \nonumber \\ & =& \left.
X_{H_0(q_0^{a(s)}(t),p_{a0}^{(s)}(t))}q_0^{(s)}(t)\right|
_{s=t}+\left. \Pi _tX_{\Pi _t^{*}H_1(q_0^{(s)})}\Pi _t^{-1}\Pi
_tq_0^{(s)}\right| _{s=t} \nonumber \\
\
& =&
(X_{H_0(q_0^{a(t)}(t),p_{a0}^{(t)}(t))}q_0^{a(t)}(t),X_{H_0(q_0^{a(t)}(t),p_{a0}^{(t)}(t))}p_{a0}^{(t)}(t))
\nonumber \\ & & \
+(X_{H_1(q_0^{a(t)}(t),p_{a0}^{(t)}(t))}q_0^{a(t)}(t),X_{H_1(q_0^{a(t)}(t),p_{a0}^{(t)}(t))}p_{a0}^{(t)}(t))
\nonumber \\
\
&=&(X_{H(q_0^{a(t)}(t),p_{a0}^{(t)}(t))}q_0^{a(t)}(t),X_{H(q_0^{a(t)}(t),p_{a0}^{(t)}(t))}p_{a0}^{(t)}(t))
\nonumber \\ \
&=&(X_{H(q^a(t),p_a(t))}q^a(t),X_{H(q^a(t),p_a(t))}p_a(t)),
\label{r13}
\end{eqnarray}
where in the fifth step, we have used the relations $\Pi
_tq_0^{(s)}=q_0^{(s)}(t)$ and $\Pi _tX_{\Pi _t^{*}H_1(q_0^{(s)})}\Pi
_t^{-1}=X_{H_1(q_0^{a(s)}(t),p_{a0}^{(s)}(t))}.$ The final line in 
(\ref{r13}%
) is nothing but the Hamiltonian equations(\ref{e0}). Therefore the
constructions(\ref{r11}) truly describes the whole evolution of measured
values of $q$ and $p.$

Now let us compare above $M_F$ description with phase space $T^{*}Q$ and
motion space $M$ descriptions. As explained in $P.21$ of the 
ref.\cite{Wood}%
, there exit changes in point of view. In the phase space $T^{*}Q$, the
system evolves by moving along an integral curve of $X_H.$ In the space of
motions $M,$ the state of system is represented by a fixed point of $M,$
while the physical observables are represented by time-dependent functions 
$%
\stackrel{\wedge }{q}^a\circ \tau _t,\stackrel{\wedge }{p}_a\circ \tau
_t,... $ on $M$ where $\tau _t:M\rightarrow T^{*}Q:q\mapsto 
(q^a(t),p_a(t)).$
In the space of free motions $M_F,$ the state of system evolves by moving
along a curve generated by interaction Hamiltonian, while observables also
are time-dependent functions $\stackrel{\wedge }{q}^a\circ \Pi 
_t,\stackrel{%
\wedge }{p}_a\circ \Pi _t,...$ on $M_{F.}$ Their actual classical
measured values at $t$ is determined by the combinational
effects. These distinctions are just analogues of distinctions
between Schr\"{o}dinger, Heisenberg and the interaction pictures
in quantum mechanics.

ZYZ thanks the IITAP of Iowa State University and Profs. J.P.Vary and
B.L.Young for their hospitality and discussions.

\end{document}